\documentstyle[epsfig,amssymb,twocolumn]{mn2e}
\newif\ifAMStwofonts

\title[Dynamical fossils in dSph galaxies]
{The survival of dynamical fossils in dwarf spheroidal galaxies in
conventional and modified dynamics }
\author[S\'{a}nchez-Salcedo \& Lora]
{F.~J.~S\'anchez-Salcedo$^{1}$\thanks{E-mail:jsanchez@astroscu.unam.mx}
and V.~Lora$^{1,2}$\\
$^{1}$Instituto de Astronom\'\i a, Universidad Nacional Aut\'onoma
de M\'exico, P.O. Box 70-264, C.P. 04510, Mexico City, Mexico\\
$^{2}$Astronomisches Rechen-Institut, Zentrum f\"ur Astronomie der
  Universit\"at Heidelberg, M\"onchhofstr.\ 12-14, 69120
  Heidelberg, Germany}

\begin{document}

\date{Accepted xxxx Month xx. Received xxxx Month xx; in original form
2009 December 10}
\pagerange{\pageref{firstpage}--\pageref{lastpage}} \pubyear{2009}
\maketitle

\label{firstpage}

\begin{abstract}
The survival of unbound density substructure against orbital
mixing imposes strong constraints
on the slope of the underlying gravitational potential and provides a new
test on modified gravities.  Here we investigate whether  
the interpretation that the stellar clump in Ursa Minor (UMi) dwarf
spheroidal galaxy is a `dynamical fossil' is consistent with Modified
Newtonian dynamics (MOND).
For UMi mass models inferred by fitting the velocity
dispersion profile, the stellar clump around the second peak of UMi
is erased very rapidly, within $1.25$ Gyr ($6.5$ orbits), 
even with the inclusion of self-gravity.
We find that the clump can hardly survive for more than $2$ Gyr
even under more generous conditions.
Alternative scenarios which could lead to a kinematically
cold clump are discussed
but, so far, none of them were found to be fully satisfactory.
Our conclusion is that the cold clump in UMi poses a challenge for both
$\Lambda$CDM and MOND.

\end{abstract}


\section{Introduction}
The standard concordance cosmological model with cold dark matter 
($\Lambda$CDM model) is remarkably successful on scales larger than
$1$ Mpc, but it faces challenges on smaller scales.
For instance, it seems that the theory predicts a too cuspy 
density profile for the dark matter at the centres of galaxies
(e.g., Trachternach et al.~2008).
The MOdified Newtonian Dynamics (MOND) proposed
by Milgrom (1983) has proven to be successful in reproducing
the kinematics of spiral galaxies without any assumption of unseen matter 
(see Sanders \& McGaugh 2002, for a review), from extremely low mass galaxies
of low surface brightness (Milgrom \& Sanders 2006) to high luminosity 
galaxies (Sanders \& Noordermeer 2007). 
Gentile et al.~(2007) found that the observed rotation 
curves in tidal dwarf galaxies are quite naturally 
explained without any free parameters within MOND, and
are inconsistent with the current $\Lambda$CDM  
theory (see also Kroupa et al.~2005).  
If MOND is able to naturally account for all of the discrepancies 
faced by $\Lambda$CDM on small scales, this would lend strong support to MOND.

In the Newtonian dark matter scenario, dwarf spheroidal galaxies 
(dSph's) require the largest mass-to-light ratios. 
Hence, dSph's provide a unique testing ground for the nature of dark matter and 
its alternatives (Gerhard \& Spergel 1992; Milgrom 1995;
Lokas et al.~2006; S\'anchez-Salcedo et al.~2006; Angus 2008).  
However, the mass-to-light ratios inferred in MOND
are very sensitive to uncertainties on the structural parameters, luminosities, 
distances and internal velocity dispersions.
It is, therefore, important to explore other gravitational effects, which, 
in principle, may offer independent tests. 

In the dark matter paradigm, the velocity dispersion profiles of the
brightest dSphs are consistent with both a cuspy NFW halo and a cored dark
halo. However,
there exists some indirect evidence that dSph galaxies may possess a core 
(Kleyna et al.~2003, hereafter K03; 
Goerdt et al.~2006; S\'anchez-Salcedo et al.~2006). 
In particular, K03 considered the survival of the cold density 
substructure detected in photometric data in Ursa Minor (UMi).
This substructure appears as
an off-centre localized stellar clump with low velocity dispersion $\simeq
0.5$ km s$^{-1}$.  
K03 concluded that the secondary peak in UMi is a long-lived structure,
surviving in phase-space because the underlying gravitational potential is
close to harmonic. In the standard dark matter paradigm, this implies that 
the dark matter halo in UMi must have a large core because, 
if the dark halo has a central density cusp, the clump should have diluted
in $\sim 1$ Gyr. 
Even if dSph galaxies are influenced to
some degree by the tidal forces exerted by the Milky Way, it is unlikely
that such a large core may have a tidal origin (e.g., Stoehr et al.~2002; 
Hayashi et al.~2003; Pe\~narrubia et al.~2008).
It is worthwhile exploring if the competing MOND scenario can explain
the survival of cold substructures in dSph galaxies in a natural way.
The question that arises is whether
the MOND gravitational potential can mimic the potential of a dark halo 
with a core
in order to explain the very longevity of the dynamical fossil in UMi.
In other words, can the inference of cored haloes in dSphs be 
accomodated naturally within MOND?

This paper is organized as follows. In section \ref{sec:tidalradius}, 
we give a general statement of the problem and outline some key
analytic results for tidal dissolution of satellite systems.
In section \ref{sec:UMidm}, we briefly 
describe the observational properties of UMi dwarf and its clump.
Some important issues regarding the assumptions and approximations made 
to study the evolution of the clump in MOND framework are given in section
\ref{sec:approximations}.
Section \ref{sec:survival} presents results on the evolution of a 
stellar clump in
MOND. In section \ref{sec:alternatives},  we discuss alternative scenarios
to account for the persistence of kinematically cold substructure in dSph 
galaxies. Finally,  our conclusions are summarized in section 
\ref{sec:conclusions}.

\section{Evolution of satellite systems and unbound clumps: 
Tidal radius and epicyclic theory}
\label{sec:tidalradius}
Suppose that a small clump of mass $M_{cl}$ is on a circular orbit
with radius $R_{\rm cir}$ around its host galaxy, which is assumed to be
spherical. 
If the circular speed with which a test particle orbits in the gravitational
potential of the host galaxy can be described by a power-law
$v_{c}^{2}\propto r^{l}$ with $-1\leq l\leq 2$,
the tidal radius is given by 
\begin{equation}
r_{t}=k \left[\frac{\chi_{\rm cir}}{2-l}\right]^{1/3} R_{\rm cir},
\label{eq:kingtidalc}
\end{equation}
where $\chi_{\rm cir}$ is the ratio between the mass of the clump and
the mass of the host galaxy mass inside $R_{\rm cir}$, and
$k$ is Keenan's (1981a,b) factor, which 
accounts for the elongation of the zero-velocity surface along the line
joining the clump and the centre of the host galaxy.
For the Keenan factor, we will take the nominal
value $k=1$ 
(see section 2 in S\'anchez-Salcedo \& Hernandez 2007 for a discussion).
Zhao (2005) and Zhao \& Tian (2006) showed that the expression for the dynamical
tidal radius given in Eq.~(\ref{eq:kingtidalc}) is also valid in MOND
as long as $\chi_{\rm cir}$ is taken as the ratio between the ``true''
baryonic masses of the satellite and the host galaxy (see S\'anchez-Salcedo
\& Hernandez 2007 for a discussion).
We see from Eq.~(\ref{eq:kingtidalc}) 
that it may happen that the clump is essentially unbound ($M_{cl}\rightarrow 0$
and $\chi_{\rm cir}\rightarrow 0$),
but it is not dissolved because $l\rightarrow 2$. 
This is true in both Newtonian and MOND gravity.
The case $l=2$ corresponds to the harmonic potential.
If the potential is quasi-harmonic,
the orbital period depends weakly on the 
energy of the stars and hence
the phase mixing is highly suppressed even if self-gravity of the
clump is ignored.

Much of early work on the dynamics of unbound satellites used the epicycle
theory to describe their evolution (e.g., Oort 1965; Wielen 1977;
Fellhauer \& Heggie 2005). 
From the linear epicyclic theory, we learn that 
it is possible to prepare a small unbound system with a nonzero
velocity dispersion that will orbit with almost no secular spreading (e.g.,
Kuhn 1993). In fact, in an idealized situation where all the stars of the clump 
have the same guiding centre, the clump would preserve its original size
in the epicyclic approximation 
because the epicycle and orbital frequencies are the same for 
all stars. 
However, even if we start with a very idealized situation in which
all the stars of the clump move on retrograde orbits resembling
epicycles with a common guiding centre,
self-gravity of the clump destroys the coherent
stellar motions and the dissolution time is infinite only when
the density of the clump is negligible. Fellhauer \& Heggie (2005)
studied the time of dissolution of such unbound stellar system, 
in Newtonian gravity, as a function of its initial internal density and
size.  They found that a system orbiting within 
a galaxy modeled as an
isothermal sphere, can survive for about $10$ galactic
orbits only if it has a very low density ($\sim 5\%$
of the background density) and small radius ($0.5\%$ of the orbital
radius). For larger internal densities of the clump,
the dissolution becomes more rapid.
In the Appendix \ref{sec:A1}, we extend the theoretical analysis of 
Fellhauer \& Heggie
(2005) to MOND gravity. We consider a small unbound clump orbiting on
a circular orbit around the centre of a galaxy.
It is shown that, because of the enhanced self-gravity in MOND, 
the dissolution time of a certain unbound clump is
shorter in MOND than in its equivalent 
Newtonian galaxy, that is, the
Newtonian galaxy (the same distribution of stars plus additional dark matter) 
which has the same ``dynamics'' as the MOND galaxy.

\section{Ursa Minor and its dynamical fossil}
\label{sec:UMidm}
\subsection{UMi: Structural parameters}
\label{sec:darkcontent}
Ursa Minor is located at a heliocentric distance of $D=76\pm 4$~kpc
(Carrera et al.~2002; Bellazzini et al.~2002).
The star formation history and the characteristics of the population
of blue straggler stars suggest that UMi is a truly `fossil' galaxy, 
where star formation
ceased completely more than $8$ Gyr ago (Carrera et al.~2002; 
Mapelli et al.~2007).
The shape of the inner isodensity contours of
the surface density of stars appears
to be elliptical with a large ellipticity ($\epsilon = 0.54$).
The King core radius of the stellar component along the
semimajor axis is $17.9'$ ($\sim 395$~pc at $76$ kpc) (Palma et al.~2003).

In the Newtonian dark matter scenario, values for $M/L_{V}$ ranging 
between $30$ to $200$
can be inferred depending on the adopted global luminosity, distance and
velocity dispersion profile. 
In particular, if we rescale the results
of Gilmore et al.~(2007) for $D=76$ kpc and use the updated luminosity
of $L_{V}=1.1\times 10^{6} L_{\odot}^{V}$ (Palma et al.~2003),
the total mass within $0.6$ kpc is $\gtrsim 7\times 10^{7} M_{\odot}$,
resulting in a $M/L_{V}\gtrsim 64 M_{\odot}/L_{\odot}^{V}$.
If we use the dispersion profile of Mu\~noz et al.~(2005), 
the $M/L_{V}$ is reduced by a factor $\sim 2$.

\begin{figure*}
\epsfig{file=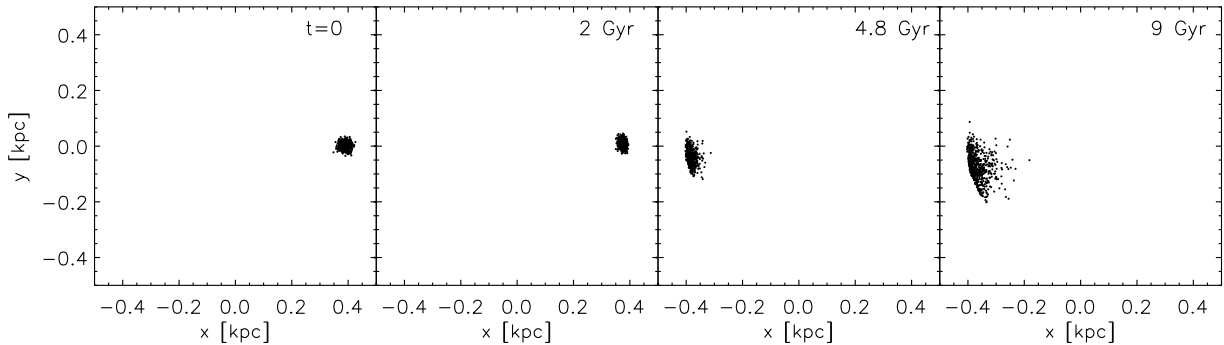,angle=0,width=17cm}
 \caption{Snapshots of an unbound clump at $t=0,$ $2$, $4.8$ and $9$ Gyr
in a dark matter halo with a scale radius $R_{s}$ equal to $2.4$ times the 
semimajor axis of the size of clump's orbit. 
The clump has an initial $1\sigma$ radius of
$12$ pc, one-dimensional velocity dispersion of $0.5$ km s$^{-1}$ and
a speed at apogalacticon equal to $0.07$ times the circular speed.}
 \label{fig:newtondm}
 \end{figure*}

\subsection{The secondary peak of UMi revisited}
\label{sec:fossil}
Several authors have reported structure in the surface density of UMi
on a variety of scales 
(Olszewski \& Aaronson 1985; Demers et al.~1995; Irwin \&
Hatzidimitriou 1995; Kleyna et al.~1998; Eskridge \& Schweitzer 2001;  
Palma et al.~2003).  
The most remarkable feature is the double off-centered peak in the 
stellar surface number density. In this paper, we concentrate on the
secondary peak, observed on the northeastern side of the major axis of UMi.

K03 found that the radius of the clump is $\gtrsim 6'$
(equivalent to $135$ pc at $D=76$ kpc); 
this is the largest circle where the isophlet contours indicate
the presence of a secondary peak in the stars.
The velocity of stars that form the secondary peak
is best fitted by a two-Gaussian population, one representing
the $8.8$ km s$^{-1}$ Gaussian background and the other 
representing a cold subpopulation of 
$0.5$ km s$^{-1}$ velocity dispersion.
The cold group of stars has the same mean velocity as the systematic
velocity of UMi, so it is likely that its orbit is close to the plane 
of the sky.

The luminosity bump can be fitted approximately by a Gaussian
density profile $\rho=\rho_{0}\exp(-r'^{2}/2r_{0}^{2})$, where
$r'$ is the radial distance to the peak in the stars,
and $r_{0}\simeq 35$ pc (e.g., Palma et al.~2003). 
The projected surface density of the clump
along its centre is: $\sqrt{2\pi}\rho_{0}r_{0}$. Since the underlying
stellar population of UMi
also contributes to the observed surface density, we need
the fraction $f$ of stars that belongs to the kinematically cold 
subpopulation.  K03 found that the subpopulation
fraction is $f=0.7$ in the best fit. 
Equating $\sqrt{2\pi}\rho_{0}r_{0}$ with $f$ times the observed surface density 
at the secondary peak,
we can infer $\rho_{0}$ and then the 
total mass in the clump $M_{cl}\simeq 16\rho_{0}r_{0}^{3}$. 
Using $f=0.7$, we obtain
\begin{equation}
M_{cl}=7.8\times 10^{4} {\rm M}_{\odot}
\left(\frac{\Upsilon_{\star}}{5.8}\right),
\label{eq:masscl}
\end{equation}
where $\Upsilon_{\star}$ is the $V$ band mass-to-light ratio of
UMi stellar population.
For a normal stellar population $\Upsilon_{\star}\approx 2$,
we have $M_{cl}=2.5\times 10^{4} M_{\odot}$.

An inspection of the stellar isophlets in UMi suggests that the secondary
peak is unbound material because of its large angular size in the plane
of the sky, as compared to the expected value for a bound cluster, and 
its bend in the isodensity contours, indicative of tidal
disruption (Palma et al.~2003). 
Let us estimate the tidal radius of a stellar cluster of mass
$M_{cl}=2.5\times 10^{4}$ M$_{\odot}$, as UMi's clump, 
on a circular orbit in a galaxy with a flat rotation curve at $R_{\rm cir}$
(i.e.~$l=0$ in Eq.~\ref{eq:kingtidalc}).  
For a UMi-like galaxy, the mass interior
to $R_{\rm cir}=390$ pc in the standard dark matter
scenario is $\simeq 10^{7}$ M$_{\odot}$. According
to Eq.~(\ref{eq:kingtidalc}), the clump 
would remain gravitationally bound only if all the 
mass that comprises the stellar cluster were placed
within a sphere of $42$ pc radius.
The appearance of such a hypothetical cluster should resemble a globular
cluster as those detected in Fornax dSph galaxy and, therefore, easily
detectable as a distinct spherical bound object. 
This is not the case for the secondary peak in UMi and, thus,
it turns out that the clump is unbound at least in the dark matter
scenario.

Since there is no evidence that the regions of density excess have
stellar populations that differ from the main body of UMi (Kleyna
et al.~1998), the most plausible interpretation 
is that this density structure, or clump, is a disrupted stellar cluster
(K03; Read et al.~2006; section \ref{sec:alternatives}). 
K03 argues that the persistence of a 
dynamically separate unbound entity
for $10-12$ Gyr, is possible only if 
the background potential is quasi-harmonic (see section \ref{sec:tidalradius}),
implying that the dark halo has a large core. 
For illustration,
Figure~\ref{fig:newtondm} shows the evolution of a clump made of
test particles,
orbiting in a spherical dark halo with a density mass profile of the form
\begin{equation}
\rho_{dm}(r)=\frac{\rho_{\rm max}}{(1+[r/R_{s}]^{2})^{1/2}},
\label{eq:densitydm}
\end{equation}
where $R_{s}$ is the scale radius.
Following K03, the central density $\rho_{\rm max}$ was selected such that
the circular velocity is $19$ km s$^{-1}$ at an angular distance of $31.5'$,
i.e.~at $\sim 700$ pc.
The contribution of the stellar mass to the gravitational potential was
ignored because its mass interior to $0.4$ kpc ($0.9\times 10^{6} M_{\odot}$)
is only $9\%$, or less,
of the dark matter mass, if we assume that UMi has a normal stellar
$V$-band mass-to-light ratio of
$\sim 2 M_{\odot}/L_{V,\odot}$.
The clump was dropped at a distance $R_{s}/2.4$ on a near radial orbit with a
tangential velocity equal to $0.07$ times the circular velocity at
its initial  position.  The clump has initially a $1\sigma$
radius of $12$ pc and one-dimensional velocity dispersion of $0.5$
km s$^{-1}$, similar to those selected by K03 in order to facilitate
comparison. The orbit lies in the $(x,y)$-plane.  From 
Figure \ref{fig:newtondm},
we see that, although the density structure becomes more
extended over time, it may survive for many Gyr if we assume
that the scale radius of the dark halo is at least $2$--$3$
times the size of clump's orbit, confirming K03 results.
Due to the phase mixing, those components of the velocity dispersion along
the orbital plane (i.e.~$\sigma_{x}$ and $\sigma_{y}$)
increase with time, whereas the perpendicular component to the orbital plane
remains constant.
As a consequence, the size of the clump does not increase
along the axis perpendicular to the orbital plane,
and remains dynamically cold in that direction
($\sigma_{z}\simeq 0.5$ km s$^{-1}$).

Lora et al.~(2009) have simulated the dynamics of a clump 
including self-gravity. 
They show that density substructure can 
persist for $\sim 12$ Gyr in a halo with a scale radius equal to 
$1.5$ times the size of the 
clump's orbit. 
Assuming that the size of the orbit is $\sim 390$ pc, a scale radius 
for the dark halo of $R_{s}=1.5 \times 390$ pc $=580$ pc is required.
Note that $R_{s}$ was defined as the radius 
at which the density has dropped a factor $\sqrt{2}$.
This large core is at odds with the predictions of $\Lambda$CDM.
Therefore, the interpretation that the clump survives
because the gravitational potential is quasi-harmonic poses a 
challenge to the standard $\Lambda$CDM, which predicts cuspy NFW profiles.
In the next sections, we investigate whether MOND can account for
the survival of the clump.

\section{The case of MOND: Concepts and approximations}
\label{sec:approximations}
In MOND, the luminous density profile determines the shape
of the gravitational potential and, thus, the index $l$ in 
Eq.~(\ref{eq:kingtidalc}).
Moreover, $\chi_{\rm cir}$ is not
sensitive to the adopted stellar mass-to-light ratio as soon as we
assume that the stellar mass-to-light ratio of the clump is similar
to that of the stellar bulk of UMi. Consequently and
according to Eq.~(\ref{eq:kingtidalc}), the tidal radius is very robust
to the adopted stellar mass-to-light ratio.
Since the only free parameter is, in principle, the stellar
mass-to-light ratio, the survival of the clump in UMi 
is an interesting dynamical test to MOND. 
We must warn here that Eq.~(\ref{eq:kingtidalc}) was derived for
MOND satellite systems embedded in {\it isolated} host galaxies.
As it will become clear later, UMi is affected by the Galactic field
and cannot be treated as it was in isolation. 

In MOND framework, the gravitational potential describing the force 
acting on a star in UMi obeys the modified Poisson equation of 
Bekenstein \& Milgrom (1984)
\begin{equation}
\bmath{\nabla}\cdot\left(\mu(x)\bmath{\nabla}\Phi\right)=4\pi G \rho,
\label{eq:bm84}
\end{equation}
where $x=|\bmath{\nabla}\Phi|/a_{0}$,
$a_{0}\simeq 1.2\times 10^{-8}$
cm s$^{-2}$ is the universal acceleration constant of the theory 
and $\mu(x)$ is 
the interpolating function which runs
smoothly from $\mu(x)=x$ at $x\ll 1$ to $\mu(x)=1$ at $x\gg 1$.
Equation (\ref{eq:bm84}) must be solved
with the boundary condition $\bmath{\nabla}\Phi\rightarrow 
-\bmath{g}_{E}$, where $\bmath{g}_{E}$ is the external gravity
acting on UMi, which has a magnitude $g_{E}=V^{2}/R_{gc}$,
where $V$ is the Galactic rotational velocity at
$R_{gc}$, which coincides with the asymptotic rotation velocity $V_{\infty}$ 
for the Milky Way. 
Although $V_{\infty}$ is very difficult to determine 
observationally in our Galaxy, 
Famaey \& Binney (2005) and S\'anchez-Salcedo \& Hernandez (2007) argue that
it must be of 
$170\pm 5$ km s$^{-1}$ by adopting a mass model for the
Milky Way under MOND. 

A clump star feels the external acceleration created by the Milky Way 
($\bmath{g}_{E}$), the acceleration generated by UMi background stars
($\bmath{g}_{I}$) and the acceleration generated
by all the other stars of the clump ($\bmath{g}_{\rm int}$).
Due to the non-linearity of the MOND field equation, the gravitational
acceleration $\bmath{g}_{I}$ 
is altered by the gravitational field of the Milky Way,
whereas $\bmath{g}_{\rm int}$ is altered by both $\bmath{g}_{E}$ and
$\bmath{g}_{I}$.

\begin{figure}
\epsfig{file=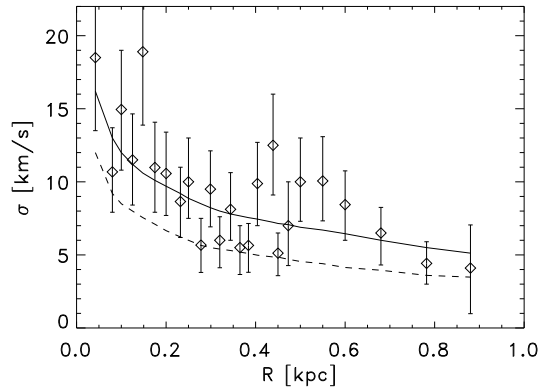,angle=0,width=8cm}
 \caption{The observed UMi line-of-sight velocity dispersion versus 
projected radius is shown as diamonds with errorbars.
The solid line corresponds to the best fitting model in MOND 
($\Upsilon_{\star}=5.8$) and the dashed line shows the predicted
velocity dispersion for $\Upsilon_{\star}=2.2$, which is at 
$1\sigma$ from the best fitting value.}
 \label{fig:veldispersion}
 \end{figure}

\subsection{Modelling UMi background gravitational potential under MOND}
\label{sec:modelling1}
Only for very special configurations (one-dimensional symmetry
--spherical, cylindrical or plane symmetric systems-- or Kuzmin discs)
the MOND acceleration $\bmath{g}$ is related to the Newtonian
acceleration, $\bmath{g}_{N}$, by the algebraic relation 
$\mu(|\bmath{g}|/a_{0})\bmath{g}=\bmath{g}_{N}$ (Brada \& Milgrom 1995).
Milgrom (1986) showed that when the mass distribution involves a small
perturbation in a medium under a uniform external field,
the perturbation in the gravitational potential 
satisfies an {\it anisotropic} Poisson-like equation
with a dilation along $z$, which is taken as the direction of the 
external field $\bmath{g}_{E}$ (see the Appendix \ref{sec:A1}).
For instance, the `modified' Plummer model for a satellite of mass $M$
in this quasi-Newtonian limit is (e.g., S\'anchez-Salcedo 2009):
\begin{equation}
\rho_{p}(\bmath{r})=\rho_{c}\left[1+\frac{x^{2}+y^{2}}{r_{p}^{2}}+
\frac{z^{2}}{(1+L_{0})r_{p}^{2}}\right]^{-5/2},
\end{equation}
\begin{equation}
\Phi_{p}(\bmath{r})=-\frac{1}{\mu_{0} }
\frac{GM}{\sqrt{(1+L_{0})(x^{2}+y^{2})+z^{2}+\bar{r}_{p}^{2}}},
\label{eq:potentialaxi}
\end{equation}
where
\begin{equation}
\rho_{c}=\frac{3M}{4\pi (1+L_{0})^{1/2} r_{p}^{3}},
\end{equation}
$r_{p}$ and $\bar{r}_{p}\equiv (1+L_{0})^{1/2} r_{p}$ are the
characteristic radii and $L_{0}$ is the logarithmic derivative of
$\mu_{0}\equiv \mu(g_{E}/a_{0})$ and hence $0\leq L_{0}\leq 1$. 
This pair is solution of the hydrostatic
Jeans equation for an isotropic velocity dispersion tensor.
We see that the potential becomes Newtonian, but with a larger
effective gravitational constant $\mu_{0}^{-1}G$, and anisotropic.
In the case of halo objects at distances $D$ much larger than the galactocentric
distance of the Sun $\sim 8$ kpc, the direction of the external field 
from the Milky Way almost coincides with the line-of-sight.

As an approximation and effective way to take into account the
external field effect (EFE) in dSph galaxies, 
Angus (2008) used the following
relation between the internal ${\bmath{g}}_{I}$ and external 
$\bmath{g}_{E}$ accelerations:
\begin{equation}
\bmath{g}_{I}\mu (x)=\bmath{g}_{N},
\label{eq:approximateEFE}
\end{equation}
where $\bmath{g}_{N}$ is the Newtonian internal acceleration of the
subsystem and $x=\sqrt{g_{I}^{2}+g_{E}^{2}}/a_{0}$. 
In view of the success of the model in reproducing the kinematics
of the classical dSph galaxies (Angus 2008), we will use 
Eq.~(\ref{eq:approximateEFE}) to derive $\bmath{g}_{I}$.
As we will see in the following, 
Equation (\ref{eq:approximateEFE}) is interpreted
as one of the simplest way to interpolate between the isolated
and the quasi-Newtonian regimes. Due to its simplicity,
it has proven to be very useful to treat the EFE (e.g.,
Wu et al.~2007; Angus \& McGaugh 2008; see also Llinares et al.~2008).
At locations where $g_{I}\gg g_{E}$, $x\simeq g_{I}/a_{0}$
and Eq.~(\ref{eq:approximateEFE}) can be simplified to
the algebraic relation $\mu(g_{I}/a_{0})\bmath{g}_{I}=\bmath{g}_{N}$,
which is the exact solution of the Bekenstein-Milgrom equation
if the subsystem is spherical or cylindrical (and it is a good approximation
in other geometries like axisymmetric exponential discs, 
e.g., Brada \& Milgrom 1995).
In the opposite regime, i.e.~in the quasi-Newtonian limit, 
where the dynamics is dominated by the
external field, that is $g_{E}\gg g_{I}$, Eq.~(\ref{eq:approximateEFE})
can be written as
$\mu(g_{E}/a_{0})\bmath{g}_{I}=\bmath{g}_{N}$.
Thus this approximation recovers the quasi-Newtonian dynamics predicted 
in linear theory, but does not capture the anisotropic dilation along
the external field direction.

\begin{table*}
\begin{minipage}{126mm}
\caption[]{A summary of relevant parameters for the reference model.
 }
\vspace{0.01cm}

\begin{tabular}{c c c c c c c c c l}\hline
{UMi} & {$D$} & {$L_{V}$} & {$r_{c}$} & {$M$} & $M/L_{V}$ &  
{$M(<r_{c})$} & { $v_{c}(r_{c})$} & $g_{E}/a_{0}$ & 
{$g_{\rm I}/g_{E}$ } \\
{}&kpc & $10^{5}$L$_{\odot,V}$ & pc & $10^{5}$M$_{\odot}$  & & $10^{5}$M$_{\odot}$ & km s$^{-1}$ &  & at $r_{c}$  \\

\hline
 & 76 & 11.0 & 395 & 63.5 & 5.8 & $26.0$ & 13.2 & 0.1 & 1.15\\
\hline
{} & {} & {} & {} & {} &  &  
{} & { } & {} \\
\hline
{Clump} & {semimajor } & {$L_{V}$} & {$R_{h}$} & {$M$} & $M/L_{V}$ &  
{$\chi_{\rm cir}$} & {$v_{c}(R_{h})$ } &  & {$g_{\rm int}/g_{E}$ }
\\
{}&axis in kpc$^{a}$ & $10^{5}$L$_{\odot,V}$ & pc & $10^{5}$M$_{\odot}$ & & &  km s$^{-1}$& & at $R_{h}$\\

\hline
   & 0.39  &  0.13 
& $55$ & 0.78 & 5.8  & $0.030$  & 4.5$^{b}$ &  & 1.0   \\
\hline
\end{tabular}
\medskip\\
$^a$ It refers to the semimajor axis of orbit's clump. 
$^b$ This is the MOND prediction (not necessarily supported by observations
--see text--).
\label{table:parameters}
\end{minipage}
\end{table*}

Suppose that we have a stellar clump embedded in
an ellipsoidal potential, such as those described by 
Eq.~(\ref{eq:potentialaxi}). 
This flattened potential creates a torque on our clump,
when the orbital plane of the clump is misaligned with the 
instantaneous direction of the external field (Wu et al.~2007).
This causes additional orbital mixing that contribute to the dissolution of
unbound clumps. Since
the approximation of Eq.~(\ref{eq:approximateEFE}) cannot capture
dilation, it neglects this source of orbital mixing.
A more refined model to include the EFE dilation
is not warranted, due to our ignorance about the density mass 
profile of UMi along the line of sight.

Angus (2008) used Jeans analysis to model the UMi velocity dispersion profile
versus circular radii (rather than elliptical radii)
reported in Mu\~noz et al.~(2005) under MOND and demonstrated that the 
best-fit model for UMi's velocity dispersion is very reasonable. 
Therefore, we will adopt exactly the same parameters for UMi as 
in Angus (2008),  which are compiled in our Table \ref{table:parameters}. 
The mass density of UMi was modelled by
a spherical King model with a core radius of $17.9'$, which corresponds
to the core radius along the semimajor axis.  The velocity anisotropy was taken
variable but the fit turned to be quite compatible with a constant anisotropy
of $0.7$.  
Figure \ref{fig:veldispersion} shows the observed velocity dispersion
from Mu\~noz et al.~(2005),
together with the best MOND fit ($\Upsilon_{\star}=5.8$, where again
$\Upsilon_{\star}$ denotes the stellar $V$-band mass-to-light ratio in
solar units), as well as the predicted MOND velocity dispersion profile 
when the $1\sigma$ error in $\Upsilon_{\star}$ is considered 
($\Upsilon_{\star}=2.2$)\footnote{The discrepant 
values for $\Upsilon_{\star}$ between Angus (2008)
and S\'anchez-Salcedo \& Hernandez (2007) is traced to the
different distance, luminosity and velocity dispersion profile used (see
section \ref{sec:darkcontent}).}.

\begin{figure}
\epsfig{file=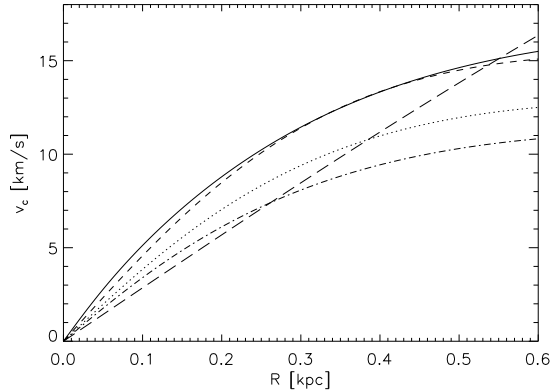,angle=0,width=8cm}
 \caption{
Circular speed curves for different models of UMi.
For the reference MOND model, which has $\Upsilon_{\star}=5.8$,
the solid line shows the MOND rotation curve, whereas the short dashed
line gives the Newtonian circular speed multiplied by a factor $2.5$ 
to show that both curves almost match.
Also shown is the rotation curve for a model with the same parameters than
the reference model but with $\Upsilon_{\star}=2.2$ (dot-dashed line).
The dotted line gives the circular speed in the reference case
($\Upsilon_{\star}=5.8$) but when the external field strength is doubled.
For comparison, the rotation
curve in Newtonian gravity with the dark matter halo described
in section \ref{sec:fossil} is also plotted (long dashed line).}
 \label{fig:rotationcurve}
 \end{figure}

In the best fitting model, the circular velocity of a test particle at the 
stellar core radius ($r_{c}=395$ pc) of UMi is $\sim 13$ km s$^{-1}$ 
(see Fig.~\ref{fig:rotationcurve}). 
Therefore, 
the characteristic internal acceleration,
$[v_{c}(r_{c})]^{2}/r_{c}\simeq 0.14\times 10^{-8}$ cm s$^{-2}$, 
is much smaller
than $a_{0}\simeq 1.2\times 10^{-8}$ cm s$^{-2}$.
On the other hand, the external acceleration, $V_{\infty}^{2}/R_{gc}
\simeq 0.12\times 10^{-8}$ cm s$^{-2}$, is also much smaller than $a_{0}$.
We conclude that UMi internal dynamics lies in the deep-MOND regime
and thus it is not sensitive to the exact form of the interpolating
function.  The ratio between the  
internal acceleration at UMi's core radius $r_{c}$ and the external acceleration
is a good measure of the importance of EFE.  
For our reference parameters given in Table \ref{table:parameters}, 
this ratio is
$\approx 1.1$, which implies that UMi dSph galaxy is at an intermediate
regime; it can be considered neither as isolated nor as external 
field dominated.

\subsection{Self-gravity of the clump}
We may define the internal acceleration $\bmath{g}_{\rm int}$
of a clump's star as 
$\bmath{g}_{\rm int}=-\bmath{\nabla}\Phi-\bmath{g}_{E}-\bmath{g}_{I}$. 
In fact, the clump is embedded in the external field
generated by all the other particles of UMi and the Milky Way. 
Due to the EFE, $\bmath{g}_{\rm int}$ depends on both 
accelerations $\bmath{g}_{E}$ and $\bmath{g}_{I}$.
Suppose for a moment that the clump was spherical and seated at
the centre of UMi. Since
$g_{I}$ is small compared to $g_{E}$
in the centre of UMi, we should recover an equation identical
to Eq.~(\ref{eq:approximateEFE}) for 
$\bmath{g}_{\rm int}$, i.e.
\begin{equation}
\bmath{g}_{\rm int}\mu (x)=\bmath{g}_{{\rm int},N},
\label{eq:approximateEFEint}
\end{equation}
where $x=\sqrt{g_{\rm int}^{2}+g_{E}^{2}}/a_{0}$ and 
$\bmath{g}_{{\rm int},N}$ is the Newtonian internal acceleration of a 
star in the clump.
If the clump is displaced at those distances from the centre of UMi 
where $g_{I}$ is not longer 
negligible relative to $g_{E}$, Equation 
(\ref{eq:approximateEFEint}) overestimates clump's self-gravity,
making it more resilient to tidal destruction.

\begin{figure*}
\epsfig{file=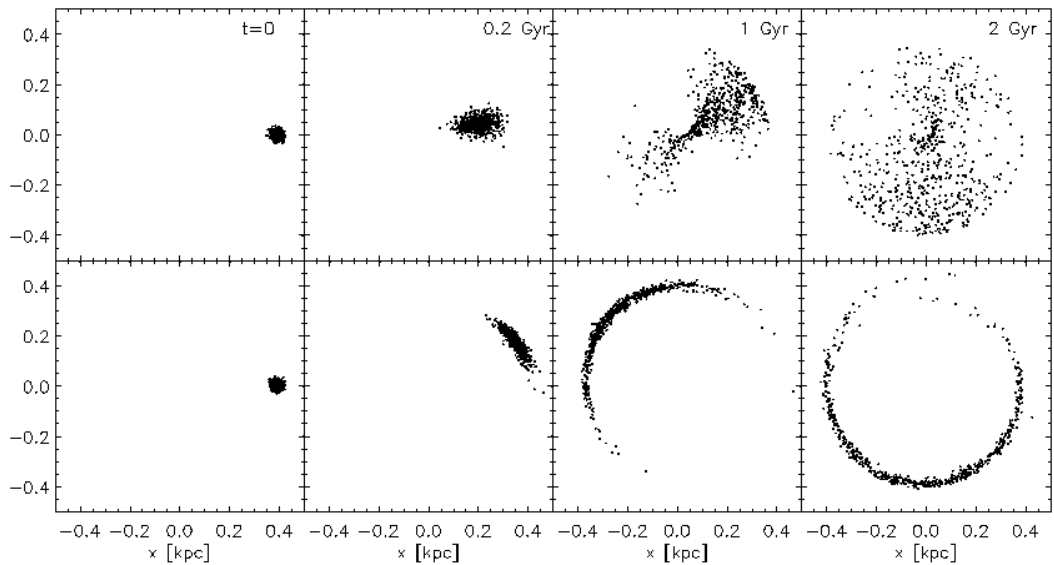,angle=0,width=15cm}
 \caption{Evolution of a clump in the reference MOND model
for a near radial orbit (top panels) and on a circular orbit (bottom panels)
when clump's self-gravity is ignored.
The orbital plane of the clump lies in the $(x,y)$ plane.
Times since the start of the simulation are indicated.}
 \label{fig:standard}
 \end{figure*}

At $g_{\rm int}\gg g_{E}$, Equation (\ref{eq:approximateEFEint}) is 
the solution of the modified Poisson equation when
our clump is spherical.
Therefore, Equation (\ref{eq:approximateEFEint}) is valid
provided that  
the mass density distribution of the clump
is close to spherical at $g_{\rm int}\gg g_{E}$. 
As time goes by, the clump and its tidal debris will form
a non-spherical mass configuration because the stars lost from the 
clump will form two tidal tails, one leading the clump and one trailing it. 
However, the break of spherical symmetry does not represent
a real limitation because, as
we will see in the next section,  
$g_{\rm int}$ is comparable to $g_{E}$ only at the
very beginning of the simulations when the clump is roughly spherical.
In fact, internal accelerations of the order of $g_{E}$
only occur at initial times and inside the clump, where the mass 
distribution can be considered as spherical.

In order to quantify the role of self-gravity, 
we will carry simulations
in which the self-gravity is turned off 
(non-selfgravitating case, NSG) and simulations 
in which self-gravity is included 
through Eq.~(\ref{eq:approximateEFEint}). 
According to the ongoing discussion, these are the limits of 
what can be expected in MOND;  the survival times derived
with Eq.~(\ref{eq:approximateEFEint}) represent upper values.
In the remainder of the paper, we will refer to 
Eq.~(\ref{eq:approximateEFEint}) as the 
reduced EFE approximation (REFE). 
Direct N-body simulations of $1200$ particles were carried out using the code
described in Lora et al.~(2009) with the REFE approximation.
The convergence of the results was tested by comparing runs with
different softening radii and number of particles.

\section{Evolution of a stellar clump}
\label{sec:survival}

\begin{figure*}
\epsfig{file=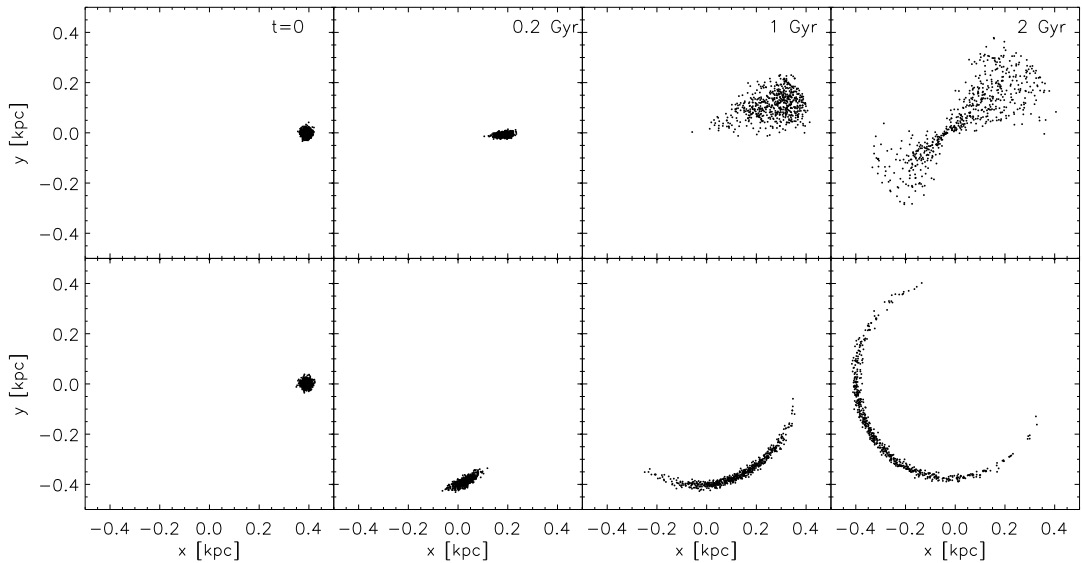,angle=0,width=15cm}
 \caption{Same as Figure \ref{fig:standard} but
for a MOND model with $\Upsilon_{\star}=2.2$. }
 \label{fig:lowvcc}
 \end{figure*}

\subsection{NSG simulations}
\label{sec:NSG}
We consider first the reference model in which $\Upsilon_{\star}=5.8$. 
Figure \ref{fig:standard} shows the disintegration of a NSG
clump on a near radial orbit with an apogalactic distance of $390$ pc
and on a circular orbit with a radius of $390$ pc.
In the case of a near radial orbit, the clump was dropped at apogalacticon
with a tangential velocity equal to $0.07$ the local circular velocity. 
The initial $1\sigma$ radius of $12$ pc and velocity dispersion of the stars in the clump of $0.5$ km s$^{-1}$ are the same as
those used in section \ref{sec:fossil}. Phase mixing dissolves the clump
very rapidly; the clump doubles its size in 
$0.2$ Gyr (essentially
one orbital revolution of the clump about UMi) and it becomes
completely diluted in less than $1.25$ Gyr ($\sim 6.5$ orbital periods).
When the clump is on a circular orbit,
the tidal debris populates the whole ring of the orbit, due to the effect
of differential rotation. 
The clump is shortlived because the  MOND circular velocity 
$v_{c}(r)$ of UMi
is essentially a scaled-version of the Newtonian (without dark matter) circular 
velocity at $r<0.6$ kpc (see Fig.~\ref{fig:rotationcurve}). 
Thus, the evolution of the clump in MOND
resembles that in a purely Newtonian galaxy with a dark halo having
a core equal to $r_{c}$, i.e.~a mass-follows-light model.
Since the core radius of the fictitious dark halo is almost equal to 
the clump's orbit, the group of stars is erased in a few crossing times. 
In other words, MOND is not able to change significantly the shape of 
the gravitational potential within the orbit of the clump, at least not 
enough to generate a near-isochrone potential needed to guarantee 
the survival of the clump.

\begin{figure}
\epsfig{file=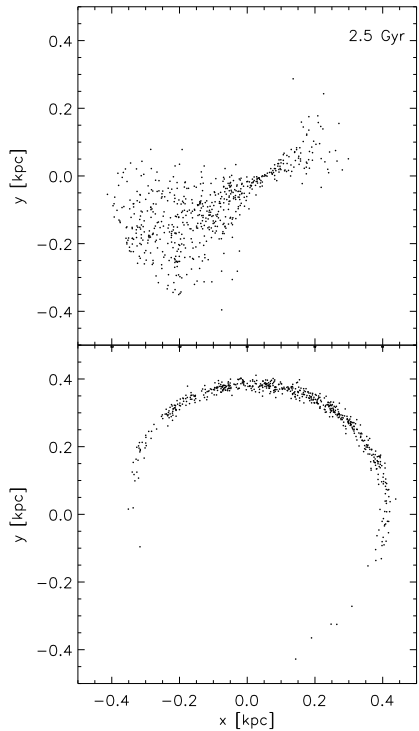,angle=0,width=10cm}
 \caption{Distribution of the stellar debris projected onto the orbital plane
at $t=2.5$ Gyr in a MOND case with $\Upsilon_{\star}=0.83$, 
when the clump with initial velocity dispersion of $0.25$ km s$^{-1}$
is on a near radial orbit (upper panel) and on a circular orbit (bottom panel).
Self-gravity of the clump was turned off.  }
 \label{fig:lowvcc_extreme}
 \end{figure}

Since the destruction time roughly scales as the galactic crossing time 
for the clump, 
one can increase the longevity of the clump by decreasing the mass of
UMi if a lower mass-to-light ratio of the stellar population of UMi
is adopted. Figure \ref{fig:lowvcc} shows the
evolution of the same clump if the body of the stars in UMi has 
$\Upsilon_{\star}=2.2$.
The clump is diluted in $< 2$ Gyr.
Even when using a $M/L_{V}$ value seven
times smaller than the preferred mass-to-light ratio 
(i.e.~$\Upsilon_{\star}=0.83 $)
and the clump is initialized with a velocity dispersion 
of $0.25$ km s$^{-1}$, the group of stars dissolves in $<2.5$ Gyr 
(see Fig.~\ref{fig:lowvcc_extreme}).
Values for the mass-to-light ratio below $0.83$ may be problematic 
because then UMi may be too fragile to the process of tidal disruption.

Similarly, the survival time can also be enhanced by assuming
a larger value for the external field acceleration $g_{E}$.
When we double the strength of the external field in our reference
model, the clump still
dissipates in $<2$ Gyr. For comparison, the circular speed for 
this model was also plotted in Fig.~\ref{fig:rotationcurve}.
The uncertainty on $V_{\infty}$ cannot 
change significantly the disruption timescale. 
Doubling $g_{E}$ corresponds to  
an asymptotic circular velocity of $250$ km s$^{-1}$, which sounds unrealistic 
in MOND.

\subsection{Simulations with self-gravity}
\label{sec:SGsimulations}

\begin{figure}
\epsfig{file=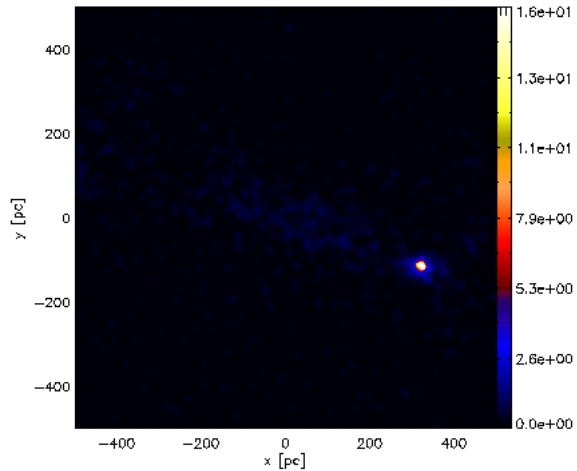,angle=0,width=8cm}
 \caption{Stellar surface density in M$_{\odot}$ pc$^{-2}$ on
the orbital plane after $6$ Gyr, for
a clump with initial $1\sigma$ radius of
$12$ pc, one-dimensional velocity dispersion of $1.8$ km s$^{-1}$ and
a speed at apogalacticon equal to $0.07$ times the circular speed.
Self-gravity of the clump was included.}
 \label{fig:standard12_6Gyr}
 \end{figure}

In the previous section, we have considered the fate
of an unbound clump in MOND.
Even if the clump is initially very compact 
(a $1\sigma$ radius of $12$ pc), orbital phase mixing 
dilutes the overdensity structure within less than $2$ Gyr in
MOND models.
When self-gravity of the clump is included, 
a clump with an initial $1\sigma$ radius of $12$ pc remains too compact 
to explain its present appearance (see Fig.~\ref{fig:standard12_6Gyr}). 
Therefore,
we have examined the disruption timescale of a clump with $1\sigma$ radius
of $35$ pc, which is similar to the present extent of the clump.
A compilation of the parameters for our reference model is given
in Table \ref{table:parameters}.
When clump's self-gravity is taken into account, a velocity dispersion
of $0.5$ km s$^{-1}$ is not enough to keep the clump in virial
equilibrium and the clump collapses to a hotter compact configuration.
In order to have the clump in a relaxed state, we needed to
adjust the initial one-dimensional velocity dispersion.

\begin{figure*}
\epsfig{file=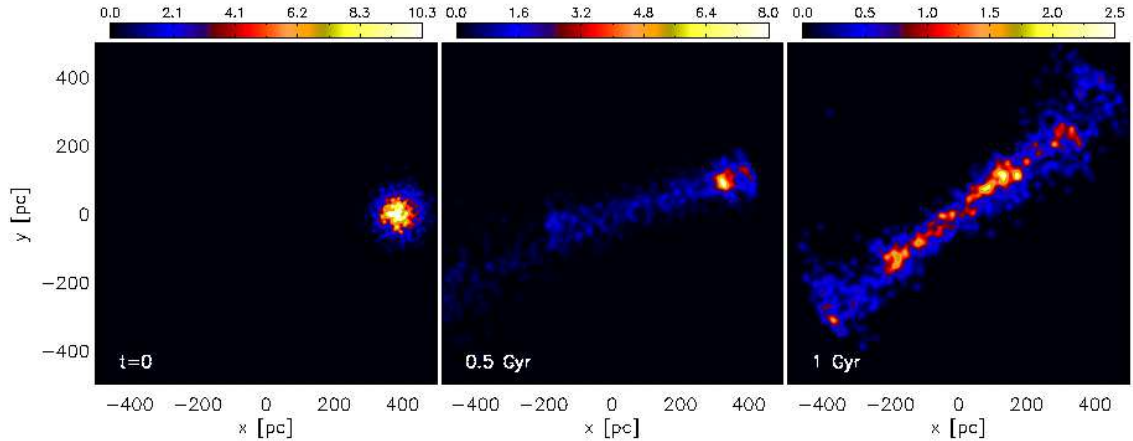,angle=0,width=16cm}
 \caption{Evolution of the stellar surface density 
in M$_{\odot}$ pc$^{-2}$ for
a clump with initial $1\sigma$ radius of
$35$ pc, one-dimensional velocity dispersion of $1$ km s$^{-1}$ and
a speed at apogalacticon equal to $0.07$ times the circular speed.
Self-gravity of the clump was included.}
 \label{fig:ecc35}
 \end{figure*}

\begin{figure*}
\epsfig{file=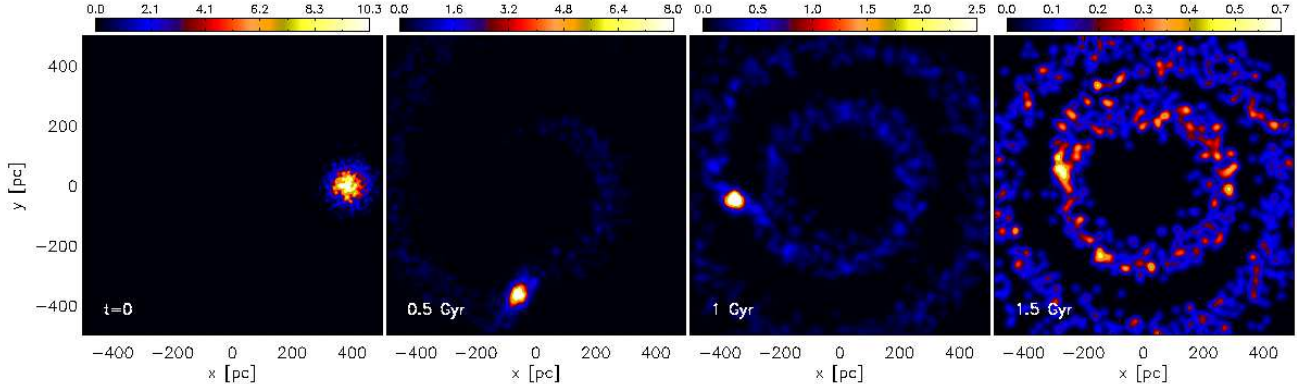,angle=0,width=18cm}
 \caption{Stellar surface density in M$_{\odot}$ pc$^{-2}$ of the
debris for a clump on a circular orbit and initial $1\sigma$ radius of
$35$ pc and one-dimensional velocity dispersion of $1$ km s$^{-1}$.
Self-gravity of the clump was included.}
 \label{fig:circ35}
 \end{figure*}

Figures \ref{fig:ecc35} and \ref{fig:circ35} show the evolution of the clump  
(assuming $\Upsilon_{\star}=5.8$)
on a near radial orbit and on circular orbit as those described in
section \ref{sec:NSG} but including self-gravity in the REFE approximation. 
The initial one-dimensional
velocity dispersion of the clump is $1$ km s$^{-1}$. Although the disruption
of the clump is slightly faster when the clump is on a near-radial
orbit, dissolution of the clump occurs within $1.25$ Gyr in both cases.
When self-gravity is turned off, 
the cluster of stars is erased within $0.5$ Gyr.

\begin{figure}
\epsfig{file=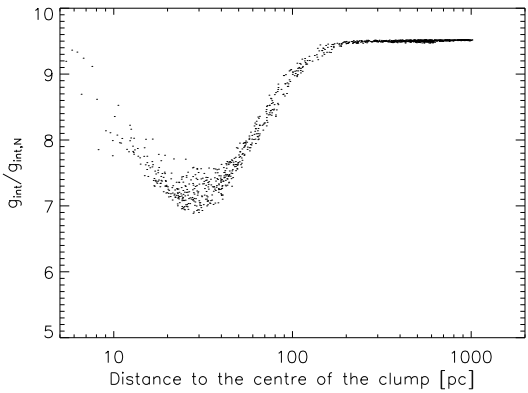,angle=0,width=8cm}
 \caption{Internal acceleration boost
for all the simulated stars as a function of the distance to the
centre of the clump, at $t=0.5$ Gyr, for the simulation shown in 
Fig.~\ref{fig:circ35}.
}
 \label{fig:boost}
 \end{figure}

In order to gain more physical insight, we will
focus on the simulation with the clump on circular orbit.
The clump cannot be considered in isolation at any time because
the mean internal acceleration due to mutual forces between
the stars of the clump, $\left<g_{\rm int}\right>$, is initially
$\simeq 0.9 g_{E}$ and decreases in time as the stars of clump 
becomes more loosely bound. For instance,  
$\left<g_{\rm int}\right>$ averaged over all the simulated 
stars is $0.37 g_{E}$ at $t=0.5$ Gyr. 
In Fig.~\ref{fig:boost}, the MOND boost factor $g_{\rm int}/g_{\rm int,N}$
at $t=0.5$ Gyr
is plotted as a function of the distance to clump's centre.
Once a star has been stripped away from the clump, at distances $>100$ pc, 
it feels a constant
boost factor and its dynamics is Newtonian but with a larger effective
Newton constant $G$. All the particles lie in the quasi-Newtonian
regime when the clump is erased.

\begin{figure}
\epsfig{file=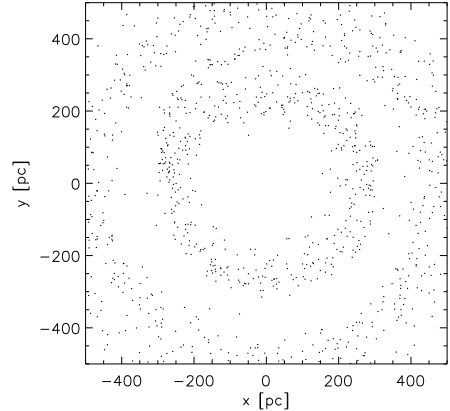,angle=0,width=6cm}
 \caption{Distribution of the simulated stars at $t=1.5$ Gyr
for the model shown in Fig.~\ref{fig:circ35}.
}
 \label{fig:circ35_ev}
 \end{figure}

Newtonian conservation laws such as the the conservation of the
total energy and total angular momentum are only preserved in
the REFE approximation when the dynamics of all the stars is quasi-Newtonian.
In order to quantify the degree of conservation in our simulations,
we have computed the net internal force $m \sum_{N} \bmath{g}_{\rm int}$
(being $m$ the particle mass)
summed over all the simulated particles and the total angular momentum.
The net internal force fluctuates in time but it is always less than
$0.5$ percent the characteristic background force $M_{cl}v_{cc}^{2}/r_{c}$,
where $v_{cc}$ is the circular speed around UMi at $r_{c}$.
The total angular momentum, on the other hand, increases by $4$ percent
during the first Gyr and by $0.5$ percent between $1$ Gyr and $1.5$ Gyr.
Therefore, the inspiraling of the clump from an initial radius of
$390$ pc to $355$ pc at $1$ Gyr (see Fig.~\ref{fig:circ35})
 is not an artifact but the result
of the gravitational transfer of angular momentum
from the clump to the trailing tidal tail.
We must also notice at this point that the density substructures displayed
in the map at $1.5$ Gyr do not have 
statistical significance, as Fig.~\ref{fig:circ35_ev} demonstrates.

A reduction of $\Upsilon_{\star}$ leads to an increase of the crossing
time of the clump around UMi's centre but 
also the mass of the clump is lowered if we assume that the clump
has the same
$\Upsilon_{\star}$ as the background stars.
For $\Upsilon_{\star}=2.2$ ($M_{cl}=3\times 10^{4}$ M$_{\odot}$),
the clump dissolves within $1.25$ Gyr as well because
clump's self-gravity is less relevant.

\begin{figure}
\epsfig{file=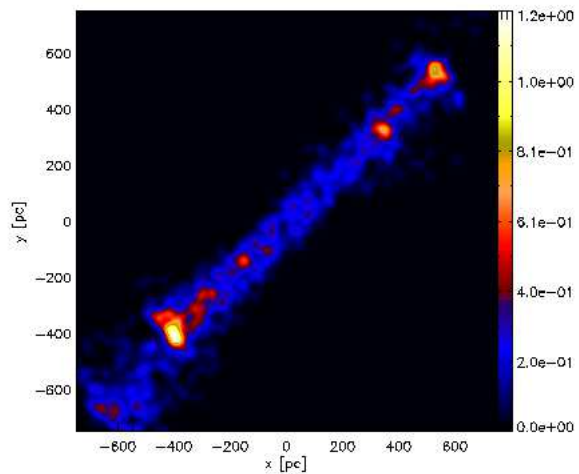,angle=0,width=8cm}
 \caption{Tidal debris of a clump with apogalacticon at $750$ pc and
on near-radial orbit, at $t=1$ Gyr.
The clump has an initial $1\sigma$ radius of
$35$ pc and one-dimensional velocity dispersion of $1$ km s$^{-1}$.}
 \label{fig:standard750_1Gyr}
 \end{figure}

It may be worthwhile exploring whether the 
destruction timescale depends on the adopted 
value of the apocenter
of the clump's orbit. For an apogalactic distance of $750$ pc, we
find that the clump is erased in $\lesssim 1$ Gyr 
(see Fig.~\ref{fig:standard750_1Gyr}).
Therefore, the persistence of the clump cannot be accounted for by
altering its orbital parameters.

Self-consistent simulations of the evolution
of a low-mass satellite were carried out by Kroupa (1997) and Klessen
\& Kroupa (1998). They found that the unbound distinguishable remnant, 
consisting of 
particles with phase-space characteristics that reduce spreading along
the orbit as those described by Kuhn (1993), 
contains a mass of about $1\%$ of the initial mass of the
progenitor. 
In the case of UMi's clump, which has a mass of 
$\sim 7.8 \times 10^{4} M_{\odot}$ for $\Upsilon_{\star}=5.8$,
the long-lived remnant should have a mass of 
$\sim 780 M_{\odot}$, too small to be detected in our simulations.

\begin{figure*}
\epsfig{file=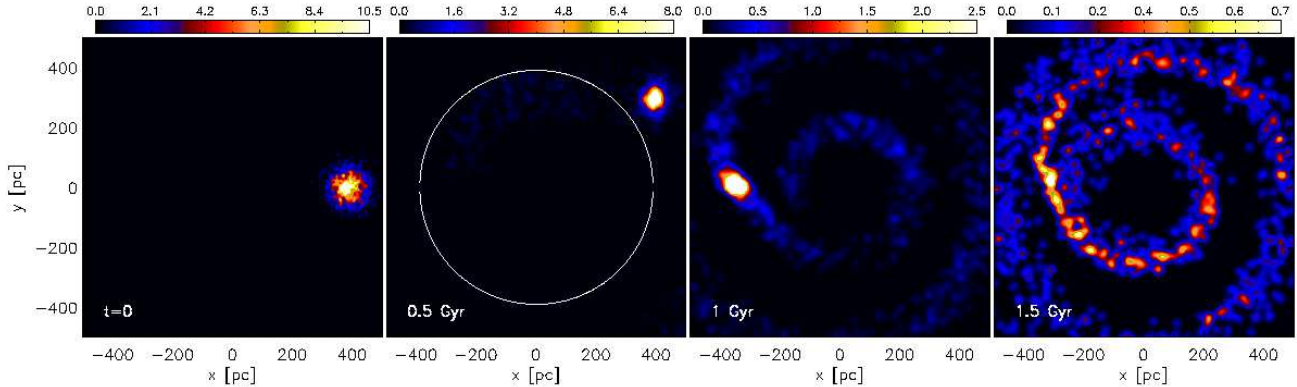,angle=0,width=18cm}
 \caption{Stellar surface density in M$_{\odot}$ pc$^{-2}$ of the
debris assuming that the orbit of UMi around the Milky Way 
has the pericentre at $40$ kpc and apocentre at $89$ kpc. 
The galactocentric distances of UMi are $R_{gc}=(76,41,81,86)$ kpc, 
at $t=(0,0.5,1.0,1.5)$ Gyr, respectively.
In this simulation, the initial clump has a $1\sigma$ radius of $35$ pc  
and starts on a circular orbit of $390$ pc radius, which
is represented by a circle in the second panel. 
Self-gravity of the clump was included.}
 \label{fig:circ35_aroundMW}
 \end{figure*}

In all our previous REFE models, the external acceleration was taken as constant
and equal to the instantaneous external acceleration of UMi at its 
current position.
The external acceleration, however, may increase with time as UMi orbit 
penetrates
deeper into the Milky Way halo when it moves towards pericentre.
Piatek et al.~(2005) measured the proper motion for UMi
and found a perigalacticon and apogalacticon of $40\pm^{36}_{30}$ kpc
and $89\pm^{71}_{11}$ kpc (at $95\%$ confidence level), respectively. 
According to Eq.~(\ref{eq:approximateEFE}),
as UMi plunges in on an eccentric orbit, it puffs up and expands 
adiabatically because $g_{I}$ gets smaller at larger external
accelerations (Brada \& Milgrom 2000).
Thus, the rotation timescale of the clump 
around UMi slows. For the same reason, 
the self-gravity of the clump lowers and the clump also inflates,
rendering itself more susceptible to tidal dissolution.  
In other words,
when the strength of the external acceleration increases, the internal
accelerations, $g_{I}$ and $g_{\rm int}$, decrease.
In Figure \ref{fig:circ35_aroundMW}, 
we show snapshots of the surface density of the
clump and its debris when UMi is on a non-circular orbit with
apogalacticon and perigalacticon at $89$ and $40$ kpc, respectively.
For our MOND value, $V_{\infty}=170$ km s$^{-1}$, the orbital period of UMi 
about the Galactic centre is $1.7$ Gyr.
In this simulation, UMi was also initially located at a galactocentric 
distance of $76$ kpc and has negative velocity, i.e., it is
approaching to the Galactic centre.
The external acceleration increases until UMi reaches pericentre after
$0.45$ Gyr. An obvious consequence of the adiabatic expansion
of UMi is that the orbit of the clump becomes more wide with
time up to $0.45$ Gyr. This can be seen clearly in the second panel
of Fig.~\ref{fig:circ35_aroundMW}; 
the orbital radius is $\sim 500$ pc at $t=0.5$ Gyr. 
By comparing Figs~\ref{fig:circ35} and \ref{fig:circ35_aroundMW}, 
we find that, at $t=1$ Gyr, the clump is able to retain more
mass in the simulation in which UMi is placed on a non-circular orbit
about the Milky Way but, still, the clump dissipates in $<1.5$ Gyr.
We can safely establish that the varying external acceleration felt by
the dwarf because of its orbital motion does not offer a promising solution to 
the survival problem of the clump.  On the contrary, Galactic tides,
not included in our simulations, may help
destroy the substructures if UMi's orbit is very elongated (say, 
perigalacticon at $<25$ kpc).

Our simulations have assumed that the mass distribution of UMi is
spherical whereas the isophlets of UMi have ellipticities of $0.54$. 
Since in MOND the gravity comes solely from the stellar component, 
the potential should be flatter, not spherical.
A perfectly spherical potential has orbits that
are confined to lie on planes. 
By contrast, orbits in axisymmetric potentials generally show
precession of their orbital planes (except for exactly planar or exactly
polar orbits), leading to an additional source of orbital mixing. 

We have explored whether a flattened mass distribution, such as that
observed in UMi, might change
the slope of the rotation curve in the equatorial plane. 
In order to estimate this effect, we have modeled the stellar bulk
of UMi as a flattened King model:
\begin{equation}
\rho(R,z)=\rho_{K}(m)\;\;\;\;\;\;\;{\rm with} \;\;\;\;
m^{2}=R^{2}+\frac{z^{2}}{\epsilon^{2}},
\end{equation}
where $\rho_{K}$ is the density profile of the spherical King model,
and derived numerically the circular velocity (see, e.g., \S 2.3 in 
Binney \& Tremaine 1987).  At radii $r<0.6$ kpc,
the UMi rotation curve in the equatorial plane is again a 
scaled-version of the curve derived in the spherical case.
If we adopt the same central volume mass density as in the spherical case,
and a flattening of $\epsilon=0.54$, the amplitude of the Newtonian rotation
curve decreases by a factor of $0.85$. Therefore, if the clump is on a 
radial orbit in the
equatorial plane of UMi, the inclusion of the ellipticity is similar to 
the use of a spherical model with $\Upsilon_{\star}=4.2$. 
Given the uncertainties in $\Upsilon_{\star}$, this is a secondary effect
and a more accurate treatment of the
flattening is not yet justified by the observations. 

We conclude that the interpretation that the stellar clump is 
a dynamical fossil is difficult to sustain in MOND because
its lifetime is less than $\sim 1.5$ Gyr even if clump's self-gravity
is taken into account. In the next section, we will discuss other
scenarios and potential problems.

\section{Alternative scenarios and caveats}
\label{sec:alternatives}
In section \ref{sec:survival}, we consider the dissolution of an unbound 
clump, treating the background as a rigid potential.  However,  
dynamical friction may induce a strong orbital decay of clusters
in dSph galaxies.
If the dark matter halo of UMi follows a NFW profile,
clusters of mass $>5\times 10^{4} M_{\odot}$ sink to the centre within
one Hubble time if they are initially
placed within $1$ kpc from the dwarf galaxy centre 
(e.g., Goerdt et al.~2006; S\'anchez-Salcedo et al.~2006; Pe\~narrubia
et al.~2009).  
In the case of MOND, the orbital decay proceeds even faster
(S\'anchez-Salcedo et al.~2006; Nipoti et al.~2008).
A fully satisfactory scenario should explain simultaneously the survival 
of the clump against phase-mixing and against orbital decay.

In the dark matter scenario, both problems are solved simultaneously
if the dark matter halo has a core (Goerdt et al.~2006; S\'anchez-Salcedo
et al.~2006). 
In MOND,  the clump can remain in orbit
as long as its starting distance is large enough so that
the dynamical friction force is highly suppressed 
due to the low stellar density of the background
(Angus \& Diaferio 2009). As already discussed, 
the clump cannot survive against tidal diffusion
even if its apocentre is placed at the tidal radius ($\sim 2$ kpc), 
because the dissolution timescale depends on the gradients of the orbital
frequencies (e.g., $d\Omega/dR$), which decay slowly with $R$.

It is important to consider other alternative scenarios for the presence 
of regions with cold density excess in dSph galaxies. 
It might be that the UMi's clump is not
a dynamical fossil but the disrupting remnant of a more massive
globular cluster in the act of dispersing.
Although this possibility requires a very fine-tuning of the initial
parameters of the globular cluster and 
circularity of its orbit, it 
might be viable if the dark matter halo is cuspy because
the density of dark matter towards the galactic centre might be sufficient 
to create a strong tidal field able to disrupt a globular cluster 
of mass $\lesssim 5\times 10^{4}M_{\odot}$
if it plunges close enough into the centre of the dwarf 
(Pe\~narrubia et al~2009).

In section \ref{sec:SGsimulations}, we saw that a compact clump
with initial $1\sigma$ radius of $12$ pc cannot be disrupted by
galactic tides. 
Hence, a typical globular cluster of mass $\sim 0.8\times 10^{5}$ M$_{\odot}$ 
is unaffected by the galactic tides in MOND,
even if the orbit brings the cluster to the centre of the dwarf galaxy. 
This is expected because the mean density of  
a typical globular cluster is higher than the mean density of a dSph galaxy.
Therefore, the clump could be a recently disrupted cluster
which survived for a long time ($\gtrsim 10$ Gyr) due to being initially
bound and {\it more masive}, 
if the parameters of the clump progenitor were fine-tuned to be initially 
much more teneous and extended than a globular cluster. 
How massive should the progenitor be in this scenario?
The cluster will be subject to 
a prolonged and slow mass-loss regime before it suffers from a sharp tidal
disruption. Pe\~narrubia et al.~(2002, 2009) and Zhao (2004) find that
cored stellar systems lose $90-95$ percent of its initial mass during
the slow mass-loss regime. This implies that the progenitor should contain
$>13$ percent of the total mass of UMi. Clearly, this model presents
severe inconsistencies because such a massive progenitor is expected to   
remain unaltered by tidal stirring. In addition, it is difficult
to explain why this cluster did not sink to the galaxy centre.

Another related possibility is that the stellar 
overdensity is not
a distinct identity but a portion of a stationary large-scale structure.
For instance, density cusps 
may emerge in the projection of eccentric discs (Tremaine 1995)
or in the projection of the tidal debris of a disrupted stellar cluster
in triaxial NFW systems (Pe\~narrubia et al.~2009).
Since simulated cold substructures always show a symmetric pattern with respect
to the dwarf centre, which has not been observed so far in the
case of UMi's clump, Pe\~narrubia et al.~(2009) suggest  
that either a significant fraction of tidal debris remains undetected
or that the origin of the clump is not related to the debris of a stellar
cluster.
In the particular case of UMi's clump, it is
unclear if, starting with a typical globular cluster, 
this model can explain simultaneously its high surface
brightness and its low internal velocity dispersion 
(Pe\~narrubia et al.~2009).
Moreover, the position of the resulting stellar substructures
in the simulations of Pe\~narrubia et al.~(2009)
are localized at a projection distance from
the galactic centre of $0.6-0.8$ times the apocentre
of progenitor's orbit. Therefore, if we assume that the orbit
of the progenitor in UMi is close to the plane of the sky, this constraint
places the apocentre of the progenitor within 
two core radius, i.e.~$\sim 800$ pc,
which might be insufficient to prevent orbital decay of the
putative progenitor by dynamical friction.

Wilkinson et al.~(2004) suggested that the cold clump in UMi is a
projection of a cold extratidal population onto the face of the dSph.
Using numerical simulations, Read et al.~(2006) explored this scenario 
in the standard Newtonian dark matter scenario and concluded that this
hypothesis is falsified for two main reasons. First, there is always a
source of hot tidal stars which mask any cold populations. Secondly,
the population of stars beyond the tidal radius always appears hot in
projection because there are more stars on circular orbits than radial
orbits at the tidal radius. Since we cannot see a reason to believe
that these arguments cannot be applied in MOND, we conclude that
if such cold clump is real, it could not have formed as a result of
tidal effects.

\section{Conclusions}
\label{sec:conclusions}

Observations of the rotation curves of spiral galaxies
strongly support a one-to-one relation between gravity at any radius
and the enclosed baryonic mass. Due to this empirical relation,
modified gravity theories like MOND are able to
account rather successfully for the amplitude and the shape of 
the rotation curves.  There are some indirect phenomena that
suggest that this empirical relation may break down at scales
of dSph galaxies (e.g., Gilmore et al.~2007, and references therein), 
challenging the interpretation of a modification
of gravity as a substitute for dark matter.

Localized regions with enhanced stellar density and, where
data permit, extremely cold kinematics have been detected in
some dSph galaxies (e.g., Olszewski \& Aaronson 1985; K03; 
Coleman et al.~2004; Walker et al.~2006). 
In the framework of CDM, K03 show that adopting a cored halo profile can
preserve the UMi's clump incorrupted for a Hubble time. This cored halo can fit
the observed stellar velocity dispersion and the persistence of the clump.
However, although we are still unable to make robust predictions about
how the dark matter distribution changes in the process of galaxy formation
when the physics of baryons are included, the formation of a large core
in a dark matter dominated galaxy, provides a hard challenge for $\Lambda$CDM. 
Mashchenko et al.~(2008) showed that energy feedback in dwarf galaxies
drives bulk gas motions and gravitational potential fluctuations 
large enough to turn the cusp into a flat core (see also Governato et al.~2009).
They find that for a Fornax sized galaxy, the dark matter core 
has an average density of $0.2 M_{\odot}$ pc$^{-3}$ 
at redshift $z=5.2$,
and the density decays a factor $\sqrt{2}$ in $\sim 300$ pc.
Still, such a core would not be enough to avoid orbital spreading of the 
clump.

In the particular case of UMi dSph galaxy, the density excess around
the secondary peak cannot be the remnant of a merger with UMi of a 
smaller, gas-rich system because the stars have the same properties
in terms of color and magnitude as the body of the UMi population
(Kleyna et al.~1998). 
One might wonder whether the density peak could instead be a projection
effect and that what we are seeing is a cold, low-density tidal tail.
Numerical experiments by Read et al.~(2006) have shown that this
scenario is very unlikely.  Another possibility is that the clump
is a portion of the stationary debris of a disrupted globular cluster.
This model is not implausible but cannot be proven yet.

A more drastic alternative scenario consists in interpreting the survival
of substructures as an internal inconsistency of $\Lambda$CDM.
We have explored if MOND can explain the longevity of unbound clumps in 
dSph galaxies.  Following K03, we have assumed that 
this overdensity structure is a disrupted stellar cluster
and simulate its evolution in the gravitational potential derived
in MOND.
Whichever the form of the orbit, the clump is
tidally disrupted within $1.5$ Gyr, even if clump's self-gravity is
included.
One can decrease UMi mass by adopting
a smaller $\Upsilon_{\star}$, slowing the process
of orbital mixing because the crossing time for the clump
becomes larger. However,  even assuming
a mass-to-light ratio of $0.8 M_{\odot}/L_{\odot}^{V}$,
the clump is disrupted in $<2.5$ Gyr. 
Our conclusion that tidal forces should have disrupted
the clump appears robust for the adopted value
of the stellar mass-to-light ratio of UMi. 

The external field acceleration felt by UMi from the Milky Way depends
on the galactocentric distance of UMi, which is time dependent if the 
orbit of UMi around the Galactic halo is elongated. 
However, the temporal variation
of the EFE can hardly boost the longevity of the clump.

In the absence of any alternative model to explain the origin of the
cold clump, we conclude that
it is challenging for both $\Lambda$CDM and MOND to explain the nature 
and dynamics of the clump in UMi.
Another alternative of gravity suggested by Moffat (2005) is 
the so-called Modified Gravity (MOG). It is a fully covariant theory
which predicts a Yukawa-like modification of Newton's law.
For a system with a baryonic mass of a few times $10^{6} M_{\odot}$ like UMi
dSph galaxy, MOG predicts little or no observable deviation from Newtonian
gravity at galactic distances of $\sim 200$ pc (e.g., Moffat \& Toth 2008).
Therefore, explaining the internal dynamics of dSph galaxies is 
problematic without advocating dark matter, weakening the appeal
of the MOG model.
Other authors achieve to find modified theories of gravity which
seem to reproduce the rotation curves of galaxies (e.g., Capozziello
et al.~2007). Nevertheless, an analysis of the dynamics of dSph galaxies 
in these theories is still missing.

\section*{acknowledgments}
We warmly thank our referee for constructive comments and
valuable suggestions that helped us improve this paper.
We are grateful to A.~Esquivel, J.~Maga\~na, A.~Rodr\'{\i}guez 
and O.~Valenzuela for useful comments.
We gratefully acknowledge support from CONACyT project CB-2006-60526
and PAPIIT IN114107.

\appendix
\section{Modified epicycles of stars in an unbound
system in MOND }
\label{sec:A1}

Fellhauer \& Heggie (2005) studied the evolution of an idealized
unbound system in a tidal field.
In this Appendix, we extend the theory to MOND. The reader is
referred to Fellhauer \& Heggie (2005) for a discussion about
the limitations of the theoretical model.

We consider a small and unbound system on a circular orbit with
radius $R_{0}$ in the axisymmetric potential
of its host galaxy. Therefore, the clump is embedded in the external 
field created by the galaxy $\Phi_{e}$. 
We use rotating, cluster-centred coordinates $\xi$,
$\zeta$ and $z$. The $\xi$-axis points to
the anticentre of the galaxy, and the $\zeta$-axis points in the
direction of orbital motion of the clump. The presence of
the clump will change the gravitational potential by some increment
$\Phi_{i}$, i.e.~$\Phi=\Phi_{e}+ \Phi_{i}$.
If $\Phi_{i}$ can be treated as a perturbation,
the gravitational field equation that governs the kinematics is
\begin{equation}
\left(\nabla^{2}+L_{0}\frac{\partial^{2}}{\partial \xi^{2}}\right)
\Phi_{i}= 4\pi \mu_{0}^{-1} G \rho
\label{eq:milgrom86}
\end{equation}
(Milgrom 1986).
Here $\rho$ is the density of the clump, 
$\mu_{0}\equiv \mu(|\bmath{\nabla}\Phi_{e}|/a_{0})$ 
and $L_{0}$ is the logarithmic
derivative of $\mu_{0}$ (in the unperturbed system).

We adopt an equilibrium model in which stars are distributed uniformly
within a triaxial ellipsoid and whose epicycles are centered at the 
centre of the ellipsoid.
The equations of stellar motion in the epicyclic approximation
(e.g., Chandrasekhar 1942) are given by
\begin{equation}
\ddot{\xi}-2\Omega \dot{\zeta}-4\Omega A \xi=-\frac{\partial \Phi_{i}}
{\partial \xi},
\label{eq:epicyclic1}
\end{equation}
\begin{equation}
\ddot{\zeta}+2\Omega \dot{\xi}=-\frac{\partial \Phi_{i}}{\partial \zeta},
\end{equation}
\begin{equation}
\ddot{z}+\nu^{2}z=-\frac{\partial \Phi_{i}}{\partial z},
\label{eq:epicyclic3}
\end{equation}
where $\xi$, $\zeta$ and $z$ are the
deviations of the stellar orbit from the circular guiding centre.
Here $\nu$ is the vertical frequency, 
$\nu^{2}=\partial^{2}\Phi_{e}/\partial z^{2}$, 
and $A$ is the Oort constant
\begin{equation}
A=-\frac{1}{2}\left(R\frac{d\Omega}{dR}\right)_{R_{0}}.
\end{equation}

Following Fellhauer \& Heggie (2005), it is possible to construct
a distribution of epicyclic amplitudes so that the space density
is uniform within a triaxial ellipsoid with semi-major axes
$a$, $\alpha a$ and $a/\Lambda_{0}$, where $a>0$ is a free parameter
that specifies the size of the clump,
$\Lambda_{0}\equiv\sqrt{1+L_{0}}$, and $\alpha$ will be determined from
the shape of closed epicycles.
The field equation (\ref{eq:milgrom86}) for $\Phi_{i}$ can be transformed 
into the standard Poisson equation
by making the substitution $\xi'=\xi/\Lambda_{0}$. 
The resultant equation can be written as
\begin{equation}
\tilde{\nabla}^{2}\Phi'_{i}=
4\pi \mu_{0}^{-1}G\rho(\Lambda_{0}\xi',\zeta,z),
\end{equation}
where $\Phi'_{i}=\Phi'_{i}(\xi',\zeta,z)$ and 
$\tilde{\nabla}^{2}=\partial^{2}/
\partial \xi^{'2}+
\partial^{2}/\partial \zeta^{2}+\partial^{2}/\partial z^{2}$.
Therefore, $\Phi'_{i}$ is the Newtonian
potential created by an ellipsoid of constant density $\rho/\mu_{0}$,
bounded by the surface $a^{2}=\Lambda_{0}^{2}(\xi'^{2}+
z^{2})+\zeta^{2}/\alpha^{2}$, which is a prolate ellipsoid with 
eccentricity $e'=\sqrt{1-(\Lambda_{0}\alpha) ^{-2}}$.
Taking advantage of the theory of homeoids, 
$\Phi'=\pi G \mu_{0}^{-1}A_{1}'\rho (\xi'^{2}+z^{2})+
\pi G\mu_{0}^{-1}\rho A_{3}' \zeta^{2}$,
where $A_{1}'=A_{1}(e')$ and
$A_{3}'=A_{3}(e')$ are given in Table 2-1 of Binney \& Tremaine (1987).
Therefore, Equations (\ref{eq:epicyclic1})-(\ref{eq:epicyclic3}) can be
written in terms of $B_{1}'=2\pi G\rho A_{1}'/(\Lambda_{0}^{2}\mu_{0})$ and 
$B_{3}'=2\pi G\rho A_{3}'/\mu_{0}$ as  
\begin{equation}
\ddot{\xi}-2\Omega \dot{\zeta}+(\kappa^{2}-4\Omega^{2}) \xi=
-B_{1}'\xi,
\end{equation}
\begin{equation}
\ddot{\zeta}+2\Omega \dot{\xi}=-B_{3}'\zeta,
\end{equation}
\begin{equation}
\ddot{z}+\nu^{2}z=-\Lambda_{0}^{2}B_{1}'z,
\end{equation}
where $\kappa$ is the epicyclic frequency if the clump itself is neglected. 
As shown by Fellhauer \& Heggie (2005), the epicycle frequency and the
vertical frequency are altered by the self-gravity of clump:
\begin{equation}
\kappa'^{2}_{\pm} = \frac{1}{2}\left(\kappa^{2}+B_{1}'+B_{3}'\pm
\sqrt{\left(\kappa^{2}+B_{1}'-B_{3}'\right)^{2}+16\Omega^{2} B_{3}'}\right),
\label{eq:newkappa}
\end{equation}
and
\begin{equation}
\nu'^{2}=\nu^{2}+\Lambda_{0}^{2}B_{1}'.
\end{equation}
If we choose the lower sign for $\kappa'$ in Eq.~(\ref{eq:newkappa}),
there exist exponentially growing solutions (i.e.~$\kappa'^{2}<0$) 
provided that 
\begin{equation}
R\frac{d\Omega^{2}}{dR}< -B_{1}'.
\end{equation}
In terms of density, one of the normal frequencies is imaginary
when $\rho$ is smaller than a certain critical density 
$\rho_{\rm cr}$,
\begin{equation}
\rho < \rho_{\rm cr}\equiv \frac{\Lambda_{0}^{2}\mu_{0}}{2\pi G A_{1}'}R\left|\frac{d\Omega^{2}}{dR}
\right|.
\end{equation}
If this condition is fulfilled, the complete solution is a linear 
combination of oscillatory solutions with frequency $\kappa'_{+}$
(those obtained by taking the upper
sign for $\kappa'$) and exponentially growing solutions:
\begin{equation}
\xi (t)=\xi_{0}\cos (\kappa'_{+}t+\psi_{0})+\lambda \exp (kt),
\end{equation}
\begin{equation}
\zeta (t)=-\alpha \xi_{0}\sin (\kappa'_{+}t+\psi_{0})-\lambda R \exp (kt),
\end{equation}
\begin{equation}
z (t)=z_{0}\cos (\nu t+\psi_{1}),
\end{equation}
where $\xi_{0}$, $z_{0}$, $\psi_{0}$, $\psi_{1}$ and $\lambda$ are free 
parameters, $k>0$ is the imaginary part of $\kappa'_{-}$,  
\begin{equation}
\alpha=\frac{2\Omega \kappa'_{+}}{\kappa'^{2}_{+}-B_{3}'},
\end{equation}
and
\begin{equation}
R=\frac{2\Omega k}{k^{2}+B_{3}'}.
\end{equation}

The equations for the epicyclic motions in the Newtonian case are naturally
recovered from the above equations just by taking $\mu_{0}=1$ and
$L_{0}=0$ (so that $\Lambda_{0}=1$). 
Due to the enhanced self-gravity of the clump by a factor
of $\mu_{0}^{-1}$ in MOND, $\rho_{\rm cr}$ 
is smaller than in the equivalent Newtonian galaxy,
i.e.~the galaxy with additional dark matter that has the same $\Omega(R)$.
In the particular case that $\rho$ is much smaller than the critical
density, we have:
\begin{equation}
k^{2}\simeq \left(\frac{4\Omega^{2}}{\kappa^{2}}-1\right) B_{3}'.
\end{equation}
We see that the dissolution time for very small initial density 
is $\propto \sqrt{B_{3}'} \propto 
\mu_{0}^{0.5} \rho^{-0.5}$ in the epicyclic approximation, which is
valid for small clumps. Since $\mu_{0}\leq 1$, the dissolution time 
is shorter in MOND than in its equivalent Newtonian galaxy.
If the gravitational potential in which the clump is embedded is harmonic, 
we have $\Omega (R)=\Omega_{0}$ and $\kappa=2\Omega$, 
then $\rho_{\rm cr}=0$ and $k=0$, implying that the dissolution time is 
infinite. This case was already discussed in in terms of the tidal radius in
section \ref{sec:tidalradius}. 

\end{document}